\documentstyle[aps,prl,multicol,epsf]{revtex}

\begin{document}
\draft

\title{Disorder-Induced Topological Defects in a $d=2$ Elastic Medium
at Zero Temperature}
\author{A. Alan Middleton}
\address{Department of Physics, Syracuse University, Syracuse, New York 13244}
\date{March 13, 1998}

\maketitle

\widetext 

\begin{abstract}
The density and correlations of topological defects are investigated
numerically in a model of a $d=2$ elastic medium subject to a
periodic quenched random potential.
The computed density of defects
decreases approximately exponentially with the defect core energy.
Comparing the defect-free ground state with the ground state with
defects, it is found that the difference is described
by string-like excitations, bounded by defect pairs, which have a fractal
dimension of $1.250 \pm 0.003$.
At zero temperature, the disorder-induced defects screen the interaction
of introduced vortex pairs.
\end{abstract}

\begin{multicols}{2}
\narrowtext

A diverse set of physical phenomena, including vortex lattices in
superconductors, incommensurate charge density waves, and crystal growth
on a disordered substrate, have
been modeled as elastic media subject to a pinning potential due to
quenched disorder \cite{PinningLengths,NarayanFisherCDW,CuleShapir}.
A possible difficulty
in applying these purely elastic models is their failure to take
into account account the possible effect of defects
in the elastic medium \cite{PinningLengths,GingrasHuse,FisherDefects}.
In general, the importance
of defects have been difficult to analyze analytically, though
recent numerical simulations and analytical arguments have indicated
that defects significantly change the $d=2$ system, but have less
effect in $d=3$
\cite{GingrasHuse,FisherDefects}.

In this paper, I present results on the effects of scalar defects on the
ground state of a model of a $d=2$ elastic medium subject to a disordered
pinning potential.
Using variants of combinatorial techniques that have been
applied \cite{gstate,topography,RiegerBlasum} to the study of
the elastic model without defects, it is possible to exactly calculate the
finite-size ground states of models that allow for defects in the
scalar displacement variable.
By comparing configurations with and without defects, 
the change in ground state due to the introduction
of defects is found to be confined to ``strings''
that connect defects.
The fractal dimension of these strings is computed to be $1.250(3)$.
The effects of the defects on the long-range response are calculated
by the introduction of a {\em fixed} pair of defects into the elastic
medium with defects.
As the system size increases, the cost of introducing the fixed pair
goes to a constant (rather than growing logarithmically), indicating
that the defects screen the long-range interaction at long
scales.

In models of $d$-dimensional
elastic media with scalar displacement subject to a pinning
potential periodic in the direction of displacement,
there are two length scales that separate distinct behaviors
\cite{PinningLengths,GingrasHuse}.
For lengths less than
the pinning length $\xi_P$, the elastic energies for deformation
are typically larger than the pinning energies, so that 
the displacement in a region of volume
$\xi_P^d$ can be represented as a single scalar variable $h$, with the
pinning energy periodic in $h$.  For length scales greater than
$\xi_P$, the pinning energy dominates and there are many metastable
states.
In a medium that allows for the creation of defects (in
the $d=2$ case considered in this paper, these defects are point
vortices), there is a length scale $\xi_V$ describing the typical
separation of vortices.  The vortices themselves are 
described by their location to within $\xi_P$
\cite{PinningLengths,GingrasHuse}.  In this work, the lattice constant
corresponds to the length scale $\xi_P$ and the distance
$\xi_V$ is controlled by a vortex core energy $E_c$.

\begin{figure}
\begin{center}\leavevmode
\epsfxsize=6.5cm
\epsffile{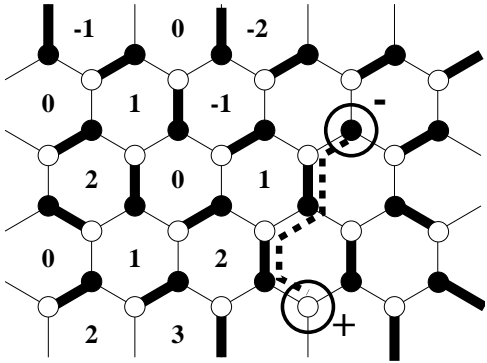}
\end{center}
\caption{
Rules for assigning heights to matchings on
a hexagonal lattice.
A- (B-) vertices are shown as
filled (unfilled) circles and matched edges
are heavy line segments.
All but the two circled vertices are matched.
Heights $h$ are defined at the centers of the hexagons.
When crossing a matched edge from an A to a
B vertex in the ccw direction about the A vertex, $h$
increases by 2; if the edge is unmatched, $h$ decreases
by 1.  A locally consistent definition of $h$ is shown on the
left side of the figure.  The right side of the figure exhibits
defects.  The unmatched vertex on the
$A$ ($B$) sublattice correspond to a negative (positive) vortex.
The dashed line indicates a possible path for a defect string that
connects two defects; if this were the defect string,
then in the ground state with no defects, the unmatched edges along
this path would become matched and the matched edges would be
removed.
}
\label{figdefects}
\end{figure}

A model for interfaces and defects in $d=2$ can be based 
upon the properties of matchings and their height representations.
Given an undirected graph $G=(V,D)$, with vertices $V$ and edges $D$,
a matching $M \subseteq D$ is a set of edges such that each vertex
belongs to at most one edge in $M$.  The graphs of interest here will be
bipartite (with subsets of vertices $A$ and $B$ and all edges
connecting $A$- and $B$-vertices),
with equal numbers of vertices in each sublattice,
and allow for complete matchings, where each vertex is
a member of exactly one edge.  In particular, consider a hexagonal
lattice with $L \times L$ two-vertex unit cells.  There exists
a one-to-one mapping between complete matchings on this
lattice, minimally frustrated states of
an antiferromagnetic Ising model on a triangular lattice,
and a solid-on-solid height representation \cite{sosmap}. 
Fig.~\ref{figdefects} indicates the mapping between matchings
and a height representation.
In the case
of a partial matching, an equal number of $A$- and $B$-vertices are
unmatched.  The uncovered vertices correspond to defects (vortices
or screw dislocations) in the
height representation.
The rules for calculating the height can still be applied
locally, but are inconsistent on any loop containing an unequal number
of $A$ and $B$ defects.
Unmatched vertices have a positive
(negative) sign, if the vertex is a member of the $B$ ($A$)
sublattice, respectively, as indicated in Fig.~\ref{figdefects}.

Consider the case of a complete matching $M$, where there is a well-defined
interface corresponding to $M$.  In the absence of quenched
disorder, the energy of the interface is independent of the height
configuration, so all matchings appear with equal weight,
and the thermally-averaged height-height correlations
$<[h(\vec r) - h(\vec 0)]^2>$ grows as
$\sim \ln(r)$ 
at all temperatures $T$ \cite{sosmap}.
In the presence of quenched disorder, this is no longer true at low
temperatures.  Theoretical calculations \cite{PinningLengths,RG}
predict that the 
2+1-dimensional model at low temperature $T$ has a correlation function
that grows as $\sim \ln^2(r)$.
This has been confirmed in the $T \rightarrow 0$ limit by numerical
calculations using the complete matching representation and other
methods
\cite{gstate,RiegerBlasum}.
The energy can be calculated in the matching
representation as $E = -\sum_{e \in M} w_e$, where $w_e$
are randomly chosen weights
associated with each edge (dimer.)  Maximizing the total weight minimizes
the energy.
When the corresponding interface in the height representation is translated
by $3$ units in the height direction, the dimer configuration is unchanged,
so that the energy is periodic in global height shifts.

In the case of an incomplete matching,
there is no longer a uniquely defined height variable corresponding
to displacements of the $d=2$ elastic medium.
An energy for the dimer configuration can then be assigned which is
consistent with (1) a local pinning energy where vertices are covered by
dimers in $M$ and (2) a defect energy for vertices that are not the
endpoints of dimers in $M$:
\begin{equation}
\label{defectenergy}
{E} = N_c E_c - \sum_{e \in M} w_e,
\end{equation}
where $N_c$ is the number of non-covered vertices and $E_c$ is the core cost
of a defect.
Minimizing $E$ in Eqn.~(\ref{defectenergy}) gives the $T=0$
configuration for the elastic medium with defects that have an associated
core energy.  Over local regions where the height is well defined, the
energy is still periodic in the height (displacement)
variable.

Ground-state configurations were found using two algorithms
that solve maximum-weight
bipartite matching (MWBM.)
The first algorithm uses a heuristic developed
by C. Zeng \cite{czprivate}, implemented in the LEDA library of algorithms
\cite{LEDA},
to directly determine the (partial) matching
that maximizes the sum in Eqn.~(\ref{defectenergy}).
The second algorithm, using almost identical processor time and less memory,
is based upon finding a maximum-weight
{\it complete} matching in a modified graph, where vertices are
duplicated and extra edges of weight $2 E_c$
and of weight zero are added to the original edge set.
A complete matching algorithm such as the cost-scaling
assignment (CSA) algorithm by Kennedy
and Goldberg \cite{CSA} can then be used to find a complete
matching on this augmented graph.
Matched vertices in the augmented graph which correspond
to edges in the original
graph give the partial matching in the original graph
that minimizes $E$.
These algorithms take \mbox{340 s} on
a 500 MHz DEC Alpha to find the ground state of a system with
$294\ 912$ vertices ($384\times384$ unit cells).

In addition, one can study the result of introducing a single
defect pair, at given locations,
into a complete matching $M$ (a defect-free medium.)
This is done by uncovering an A- and a B-vertex and
arranging $|M|-1$ dimers to cover the
remaining vertices.  This arrangement can be found by
solving a shortest paths problem
$p$ in a directed graph $G'$ which is determined by $M$ and $G$.
An undirected edge $e \in G$ is 
replaced with a directed edge $e' \in G'$ from the A- to a B-vertex
if $e \in M$, otherwise $e'$ is directed
from the B- to the A-node. A directed path $p$ following the
edges in $G'$ is assigned the energy change
$\Delta {E}(p) = \sum_{e \in p} w'_e$ where $w'_e = w_e$ if $e \in M$ and
$w'_e = -w_e$ for $e \not\in M$.
Minimal energy change
paths starting from an introduced positive defect to any other vertex
then give the minimal energy
excitation due to a defect pair introduced at the endpoints of the
path.
Such paths can be determined by a
shortest path (SP) algorithm that allows for negative weights;
for this purpose, the Goldberg-Razik algorithm was used
\cite{GoldbergRazik}.
Note that this algorithm is distinct from the shortest paths
algorithms used to study
the directed polymer problem \cite{KardarZhang}, where the edge weights can
all be made positive by a uniform shift without affecting
the paths.

Finally, the combination of the introduction of a fixed pair and defects with
a chosen
core energy was studied using a MWBM algorithm on a graph
$G'' = (V, D \cup \left\{z\right\})$ with $z$ an external
edge connecting an A-node and a B-node at a separation
${\vec{r}}_{def} = (L/2,L/2)$ (in the lattice unit vector representation),
with $w_z = \infty$, so that $z$ always
introduces a pair of defects separated as far as possible
in a finite system.  The MWBM algorithm then gives the minimum $E$
for the introduced defects, in the presence of disorder-induced
defects.

Simulations using MWBM without an introduced defect
were performed for a variety of system sizes and defect core
energies.  The edge weights (pinning energies) were chosen
from a uniform distribution
in the range $[0, 1)$, with a discrete resolution of $10^{-4}$.
Defect densities were computed by comparing $|M|$ with 
the number of edges in a complete
matching (a decrease in dimer number of one gives a defect pair.)
Complete matching
($E_c = \infty$) configurations were
compared with the partial matchings for identical
disorder to determine the changes
due to defects.
For the
largest system size ($384 \times 384$ unit cells), 400 samples were
studied.
Simulations for introduced defect pairs using the
SP algorithm were performed for systems up to $768 \times 768$ unit cells
($> 4\ 360$ samples.)

In the MWBM calculation with finite core energy, defects are readily
identified as unmatched vertices ($+$ if a B-vertex, $-$ if an A-vertex.)
The total defect density $\rho = (N_+ + N_-)/2WL$ is fit to within
statistical errors by a simple exponential,
$\rho \propto \exp(-E_c/E_0)$, with $E_0 = 0.45$, 
for $\rho^{-1/2} > 10$.  This form is consistent with
the minimum core energy $E_c^{\min}(L)$
that typically excludes {\it introduced} defects on a length
scale of $L$; numerical calculation of this quantity using the
SP algorithm on complete matchings gives
$E_c^{\min}(L) \approx ({\rm const.}) + (0.36) \ln(L)$.

\begin{figure}
\begin{center}
\leavevmode
\epsfxsize=6.5cm
\epsffile{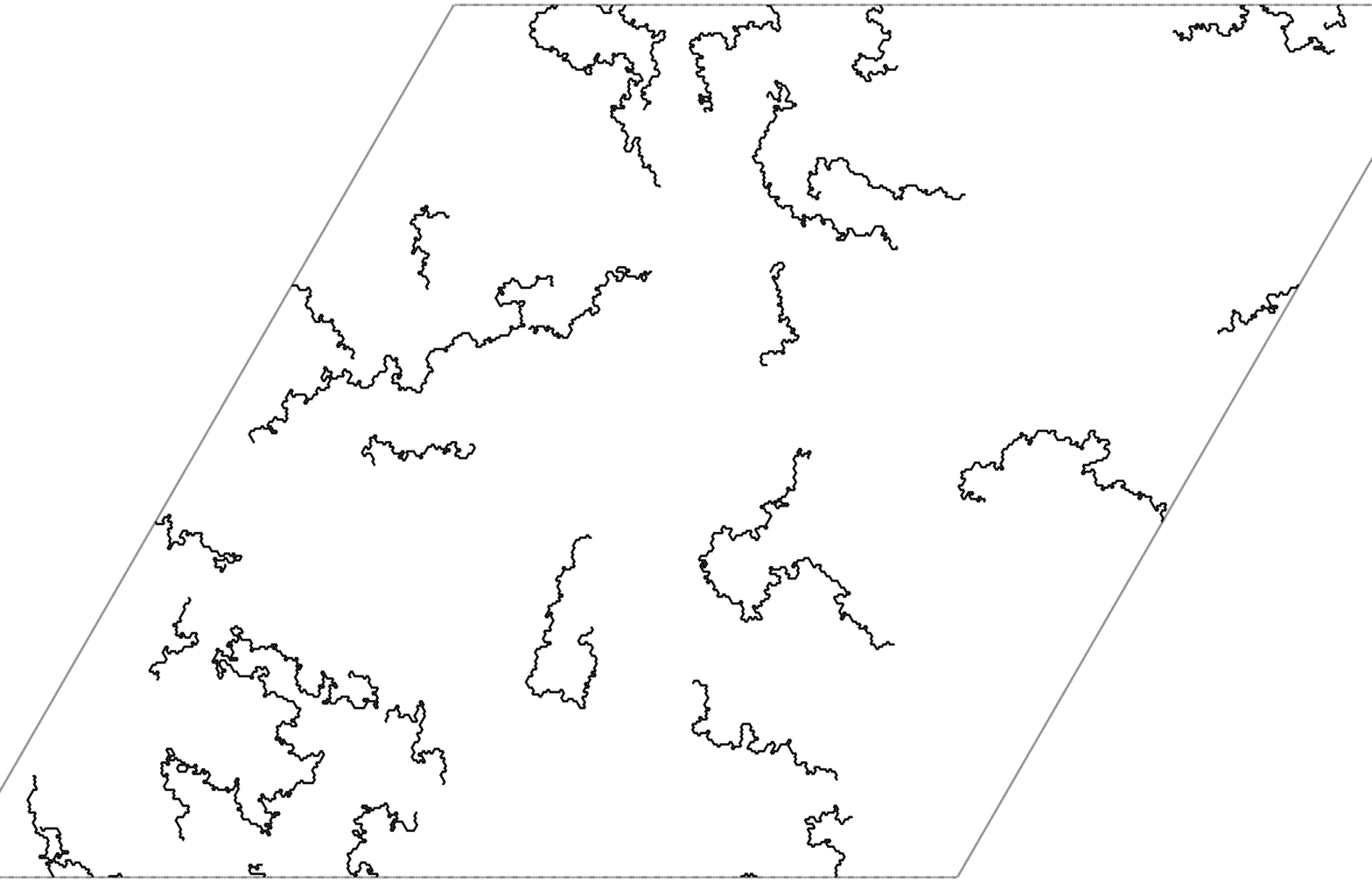}
\end{center}
\caption{
The symmetric difference between matchings for a ground state
with no defects and
a ground state for the same disorder realization, with a
defect energy of $E_c = 1.2$ ($384\times384$ unit cells.)
Dimers are included if they belong to a matching in one of the ground
states, but not both.
Vortices are at the ends of the defect lines.  The defect lines
themselves show where the phase change due to the 
introduction of defects is localized.
}
\label{figdefectspic}
\end{figure}

The ground states with and without defects can be directly compared.
There are two types of changes apparent: the introduction of vortices
and changes in the matching in the regions between vortices.
The changes are realized as
``strings'' connecting pairs of defects, as shown in Fig.~\ref{figdefectspic}.
This result is consistent with the work
of Gingras and Huse \cite{GingrasHuse} who argued that
in an $XY$ model, the phase difference caused by a defect will be
confined to lines connecting vortices of opposite sign.
This confinement results from strong pinning on scales larger than
$\xi_P$; the height difference of $\pm 3$ found on a loop enclosing
a vortex is not spread out uniformly, but takes place over the scale $\xi_P$. 

\begin{figure}
\begin{center}
\leavevmode
\epsfxsize=8.0cm
\epsffile{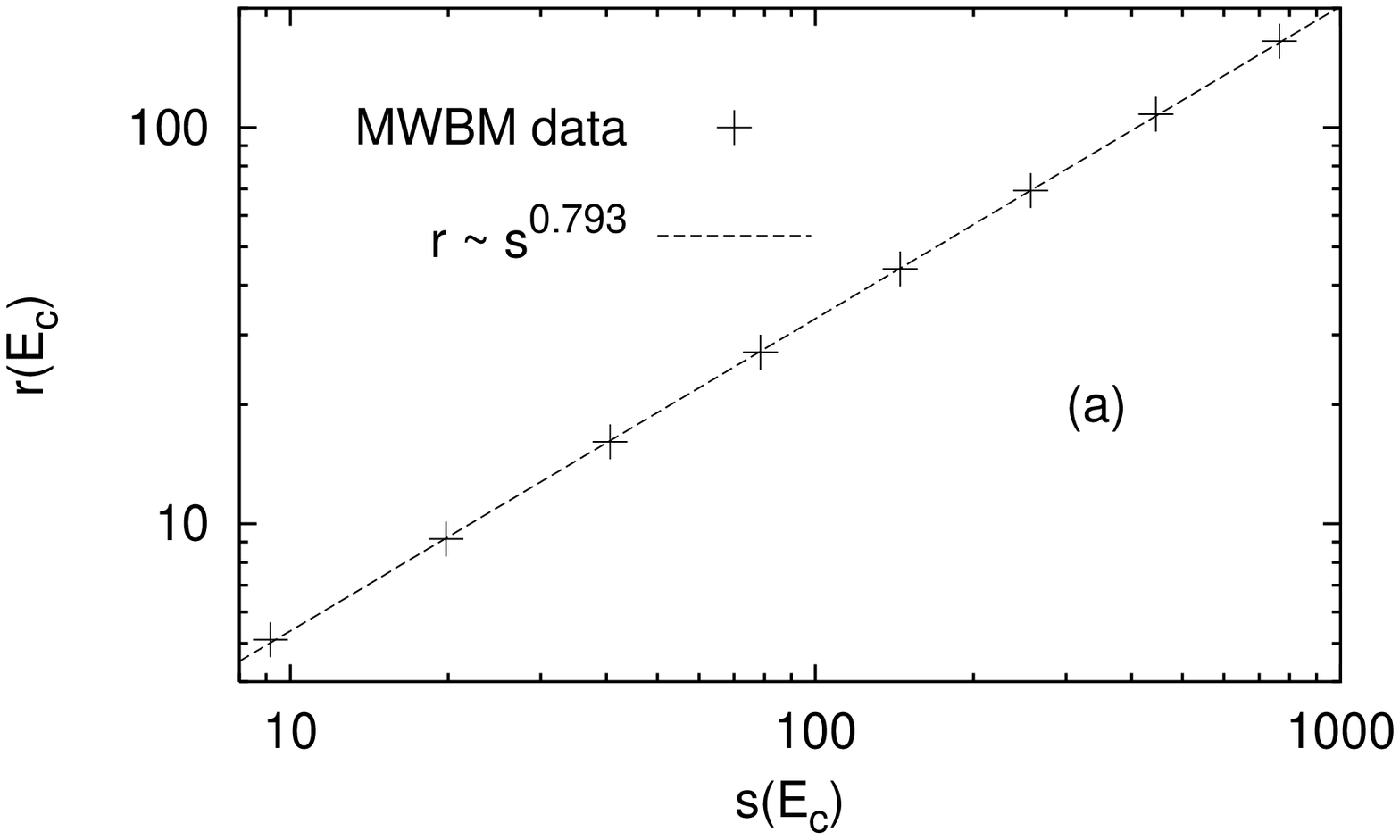}
\end{center}
\begin{center}
\leavevmode
\epsfxsize=8.0cm
\epsffile{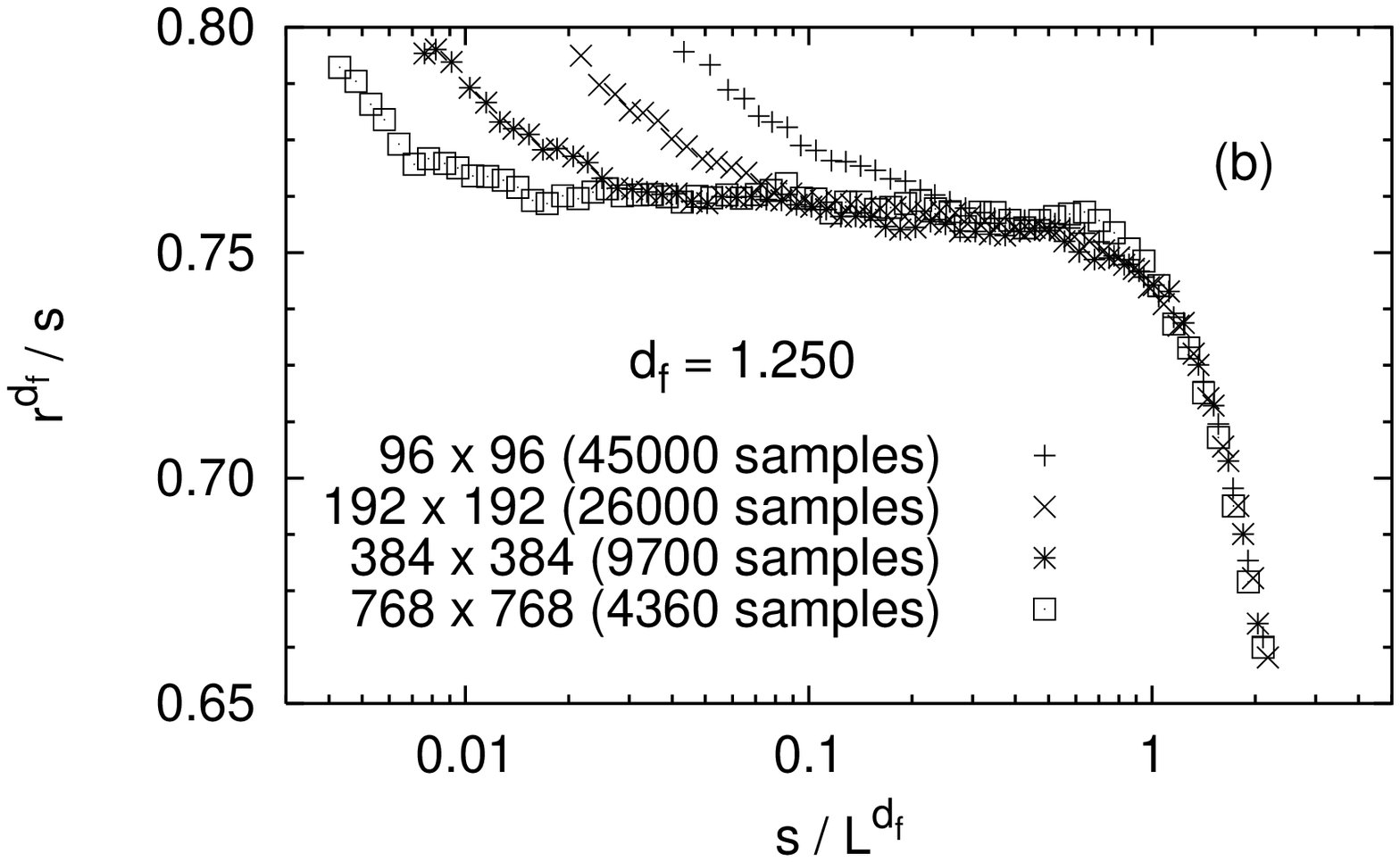}
\end{center}
\caption{
(a) A plot of
the mean end-to-end
displacement $r(E_c)$ of defect strings
vs.
the mean length
$s(E_c)$ of a defect path,
for core energies $E_c=0,0.2,0.4,\ldots,1.4$, found by examining
defect strings as shown in Fig.~\ref{figdefectspic}.
Statistical and systematic
(from the extrapolation $L\rightarrow\infty$) uncertainties are less than
1\% in each quantity.
The straight line indicates a fit to the power law $r \sim s^{1/d_f}$,
for $d_f = 1.261(16)$.
(b) A finite-size scaling plot of $r^{d_f}/s$
as a function of $s/L^d_f$ from
spanning trees generated by the SP algorithm on complete matchings.
In this case, precise collapse over two decades with a linear
abscissa gives the fitted value of $d_f$ as $1.250(3)$.
}
\end{figure}

The strings corresponding to the difference between ground states with
and without defects
were characterized by the string length
$s$ (number of edges in the string) and the end-to-end distance $r$.
The mean of the
end-to-end displacement distance of the paths,
$r(E_c)$, and the mean length of the paths, $s(E_c)$, depends on the
core energy.
The dimension $d_f$ was computed assuming
\begin{equation}
r(E_c) \sim s^{1/d_f}(E_c).
\end{equation}
The quantities $r(L, E_c)$ and $s(L, E_c)$ were simply
computed by averaging over all defect paths at fixed $L$ and
$E_c$.
The limits $s(L\rightarrow\infty, E_c)=s(E_c)$ and
$r(L\rightarrow\infty, E_c)=r(E_c)$
were found by scaling fits.  For example,  a
plot of $r(L, E_c)/r(\infty, E_c)$ vs.\ $L/r(\infty, E_c)$ was
made for trial values
of $r(\infty, E_c)$ until the data collapsed to a single line.
The values of $s(L, E_c)$ were scaled using a presumed
scaling form
$s(L, E_c)/s(\infty, E_c)
\sim L / s(\infty, E_c)^{1/d_f}$, for trial values of 
$s(\infty, E_c)$ and $d_f$.  The final value of $d_f$ was
then most accurately read from a plot of 
$r(\infty, E_c)$ vs. $s(\infty, E_c)$, as shown
in Fig.~3(a).
The single defect paths caused
by the two vortices placed in a complete matching, computed
using the CSA and
SP algorithms, were also studied.
In a single sample, $s$ can be calculated for all possible values of $r$;
the finite-size scaling is shown in Fig.~3(b).
The fractal dimensions for the multiple defect and
single introduced defect, $d_f = 1.261(16), 1.250(3)$,
respectively, are numerically consistent.

\begin{figure}
\begin{center}\leavevmode
\epsfxsize=7.0cm
\epsffile{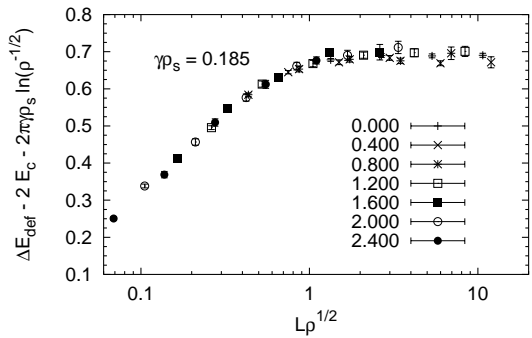}
\end{center}
\caption{
A plot of the energy cost $\Delta {E}_c$
to introduce two maximally separated defects
into a system of size $L \times L$, scaled to indicate consistency with
the limits described in the text.  The symbols indicate
the core energy $E_c$; distinct points with the same symbol are found
by varying $L$ ($L = 6,\ldots, 192$.)
Error bars are statistical ($1\sigma$.)
The coefficient $\gamma\rho_s$ was chosen for the best scaling collapse.
For fixed $E_c$, the defect cost approaches a
constant in large systems, indicating the screening of the defect-defect
interaction.
}
\label{defectcost}
\end{figure}

The SP algorithm gives the minimum energy excitations for a pair of
defects at all separations, in the absence of other defects.  Applying
MWBM to the extended graph $G''$ allows one to study a pair of defects
at fixed separation in the presence of defects.  The cost
$\Delta {E}_{def}$
to introduce a defect pair is given by the difference between
the ground state
energies for $G$ and $G''$ at the same core energy.
Introducing a pair of defects, if other defects can be ignored, costs an
energy $2 E_c$ (for the defect cores) and a
separation energy,
logarithmic in the separation \cite{FisherDefects}, so that at small values of $x = \rho^{1/2} L$,
$\Delta {E}_{def} = 2 E_c + c + 2 \pi \gamma \rho_s \ln(L)$,
where $\rho_s$ is the long-distance
stiffness associated with a single core, $c$ is a constant, and
$\gamma$ is a constant given by the sample geometry.
For large values of $x$, the introduced pair cost approaches
a constant, if the disorder-induced
defects screeen the interaction between the introduced
defects, with the crossover occuring for $\rho^{1/2} L \sim 1$
(giving
$\Delta {E}_{def} = 2 E_c + c'
+ 2 \pi \gamma \rho_s \ln(\rho^{-1/2})$ for
large $x$.)
The results of the simulation, as plotted in Fig. (\ref{defectcost}), are
consistent with this form of defect screening.

By applying optimization algorithms to many large samples,
the lack of long-range elastic behavior in a disordered
medium at $T=0$ with defects
is confirmed and the string dimension $d_f$ is precisely
determined.
Note that $d_f$ appears to be distinct from the defect-step dimension
$1.35 \pm 0.02$ previously reported \cite{RiegerBlasum}; this difference
is surprising and further confirmation would be useful.
The dimension $d_f$ is very close to that
of 5/4 for loop-erased random walks (LERW's) \cite{Majumdar},
and distinct from the dimension
(1.22(1) \cite{cieplaketal}; 1.222(3) \cite{mstaam})
of minimal-spanning trees (if the weights were all positive in $G''$,
the search for defect
strings would give a minimal spanning tree; note that the SP algorithm
has a loop-cancellation similar to that of the LERW.)

I thank Chen Zeng and Daniel Fisher for stimulating
discussions of their
unpublished work.
This work was supported by
the National Science
Foundation (DMR-9702242) and by
the Alfred P. Sloan Foundation.

\end{multicols}
\end{document}